\documentclass[preprint, aps,showpacs]{revtex4}

\begin{document}
\bibliographystyle{apsrev}
\title{Luttinger Parameter $g$ for Metallic Carbon Nanotubes and Related Systems}

\author{William Que}
\email[]{wque@ryerson.ca}
\affiliation{Department of Mathematics, Physics and Computer Science, Ryerson University, 350 Victoria Street, Toronto, Ontario, Canada M5B 2K3}

\date{  2002}

\begin{abstract}
The random phase approximation (RPA) theory is used to derive the Luttinger parameter g for metallic carbon nanotubes. The results are consistent with the Tomonaga-Luttinger models. All metallic carbon nanotubes, regardless if they are armchair tubes, zigzag tubes, or chiral tubes, should have the same Luttinger parameter g. However, a (10,10) carbon peapod should have a smaller g value than a (10,10) carbon nanotube. Changing the Fermi level by applying a gate voltage has only a second order effect on the g value. RPA theory is a valid approach to calculate plasmon energy in carbon nanotube systems, regardless if the ground state is a Luttinger liquid or Fermi liquid.  
(This paper was published in PRB 66, 193405 (2002). However, Eqs. (6), (9), and (19) were misprinted there.) 
\end{abstract}
 
\pacs{71.10.Pm, 78.67.Ch, 73.20.Mf}

\maketitle

In the past decade, the study of Luttinger liquids$^1$ has been fuelled by experimental advances in one-dimensional (1D) systems such as semiconductor quantum wires,$^2$ and more recently, carbon nanotubes.$^3$  The non-Fermi liquid behavior of an interacting 1D electron gas is predicted by Tomonaga-Luttinger models.$^4$ The assumed linear single-particle energy spectra render these models exactly solvable by bosonization method, yielding eigen modes as collective charge density modes (plasmons) and spin density modes. Single-particle excitations are forbidden in the Luttinger liquid. For wavevector $q>0$, the single-particle excitation energy (for the non-interacting electron gas) in these models is linear in $q$, $E_s=q\hbar v_F$, where $v_F$ is the Fermi velocity, while the collective charge density mode has energy
\begin{equation}
E_c=q\hbar v_F\sqrt{1+{2V_q\over \pi \hbar v_F}}.\label{E1}
\end{equation}
Here $V_q$ is the eletron-electron interaction matrix element, and typically the value at $q=0$ is taken. The Luttinger parameter $g$ is defined as the ratio of the two energies,
\begin{equation}
g=E_s/ E_c=1/\sqrt{1+{2V_q\over \pi \hbar v_F}}. \label{E2}
\end{equation}
This parameter is of key importance in the study of Luttinger liquids. It uniquely determines the exponents $\alpha$ of the power law behavior of the conductance with respect to temperature and voltage, and nonzero $\alpha$ values for carbon nanotubes are indications of a Luttinger liquid.$^{5-9}$

In 1992, Li, Das Sarma, and Joynt$^{10}$ noted that the standard random phase approximation (RPA) theory applied to a 1D semiconductor quantum wire produced results in quantitative agreement with the plasmon dispersion measured by Go\~ni {\it et al.}$^2$ They also obtained an analytical solution for the plasmon energy in the case where only one subband is occupied by electrons. If the small $q$ limit is taken, the analytical solution reduces to the result in Eq. (1). The agreement of the RPA theory with the exact result for Tomonaga-Luttinger models in the small $q$ limit dispelled the myth that RPA was not suitable in 1D. The need to take the small $q$ limit in order to obtain agreement is due to the fact that in Ref. 10, a parabolic single-particle energy spectrum is used. Near the Fermi surface, the single-particle excitation energy for a non-interacting electron gas is 
\begin{equation}
E_s={\hbar^2(k_F+q)^2\over 2m^*}- {(\hbar k_F)^2\over 2m^*}=q\hbar v_F+ {(\hbar q)^2\over 2m^*}, \label{E3}
\end{equation}
where the Fermi wavevector $k_F$ is related to the Fermi velocity $v_F$ by $v_F=\hbar k_F/m^*$. Only when $q$ is small, is the last term in Eq. (3) negligible and the linear energy spectra in the Tomonaga-Luttinger models become good approximations. In real systems, the single-particle excitation energy is usually nonlinear. Hence the bosonization method and the RPA theory each has its advantages and shortcomings: The bosonization method is robust and exact for linear single-particle energy spectra, but is applicable only in the small $q$ region where the dispersion is linear; The RPA theory is approximate in general but its validity is not limited to the small $q$ limit. 
In RPA, the plasmon energy is determined by the zero of the dielectric function
\begin{equation}
\varepsilon (q,\omega )=1-V_q P(q,\omega)=0, \label{E4}
\end{equation}
where $ P(q,\omega)$ is the polarization of the non-interacting electron system,
\begin{equation}
 P(q,\omega)=\sum_{i} P_i(q,\omega)={2\over L}\sum_{i,k}{f_{i,k+q}-f_{i,k}\over E_{i,k+q}-E_{i,k}-\hbar\omega}.\label{E5}
\end{equation}
In the above equation, $L$ is the length of the 1D system, the factor of 2 comes 
from spin degeneracy, $i$ is the subband index, $ E_{i,k}$ is the single-particle energy, and $ f_{i,k}$ is the Fermi distribution function. 

In this paper we derive the Luttinger parameter $g$ for metallic carbon nanotubes. As is well known, the structure of carbon nanotubes is specified by an integer index pair $(n,m)$. Those with $m=n$ are called armchair carbon nanotubes, those with $m=0$ are called zigzag carbon nanotubes, and others are called chiral carbon nanotubes.$^{11-14}$ All armchair carbon nanotubes are metallic, zigzag tubes with n being multiples of 3 are metallic, and chiral tubes with $2n+m$ being a multiple of 3 are metallic. 
Unlike in Ref. 10 where there is only one occupied subband and the single-particle energy dispersion is parabolic, carbon nanotubes have more (partially) occupied subbands and have sinusoidal band structures.$^{11-15}$ 
The band structure of carbon nanotubes can be easily derived from that of the graphite sheet, given by$^{11-13}$  
\begin{equation}
E_{graphite}(k_x,k_y)=\pm \gamma_0 [1+4cos({\sqrt{3}k_xa\over 2})cos({k_ya\over 2})+4cos^2({k_ya\over 2})]^{1/2}, \label{E6}
\end{equation}
where $a=0.246$ nm  is the lattice constant of a graphite sheet, and $\gamma_0$ is the overlap integral between nearest-neighbor carbon atoms. By rolling up a graphite sheet along different directions, one obtains differently structured carbon nanotubes.

The Brillouin zone of the graphite sheet is hexagonal, and $k_x$ and $k_y$ can be any value within the first Brillouin zone. Once the graphite sheet is rolled up into a nanotube, however, allowed values for $k_x$ and $k_y$ are reduced due to quantization in the circumferential direction.  
For the armchair carbon nanotube, $k_x$ is in the circumferential direction and can only take on quantized values $k_x=\nu 2\pi/n\sqrt{3}a$, with $\nu=0,1,2,..,n.$ Most of the subbands in an armchair carbon nanotube are either fully below or fully above the Fermi energy $E_F$, with only two subbands crosssing the Fermi level (those obtained when $\nu=n$).$^{11-15}$   
For low energy (intrasubband) excitations, only these two subbands need to be considered. This is not an approximation, because other fully occupied or fully empty subbands only contribute to intersubband excitations, which have different angular momenta from intrasubband excitations. (In the carbon nanotube, the angular momentum is a good quantum number, therefore intersubband and intrasubband excitations are decoupled.)
The energy dispersions of the two subbands crosssing the Fermi level are
\begin{equation}
E_{i,k}=\pm \gamma_0[2cos(k_ya/2)-1].\label{E7}
\end{equation}
In Eq. (7) we take the convention that  $i=$1 ($i=2$) corresponds to the $-$ ($+$) sign. The Fermi wavevector is determined by letting $ E_{i,k}=E_F=0$, hence $k_F=2\pi/3a$. 
Figure 1 shows the dispersions of the two subbands.

Near the K points where the two subbands cross the Fermi level and each other, the single-particle excitation energies are 
\begin{equation}
E_s=E_{i,k_F+q}- E_{i,k_F}=\pm 2\gamma_0[cos{k_F+q\over 2}a-cos{k_F\over 2}a]=\mp 4\gamma_0sin{qa\over 4}sin({k_Fa\over 2}+{qa\over 4}).\label{E8}
\end{equation}
In the small $q$ limit, to first order in $q$, Eq. (8) reduces to
\begin{equation}
E_s=\mp q\gamma_0a sin({k_Fa\over 2})=\mp q{\sqrt{3}\over 2}\gamma_0a.\label{E9}
\end{equation}
If we compare this to the linear energy spectra of the Tomonaga-Luttinger models $E_s=q\hbar v_F$, we see that in the case of armchair carbon nanotubes, we should equate 
\begin{equation}
\hbar v_F=\gamma_0a sin({k_Fa\over 2})={\sqrt{3}\over 2}\gamma_0a.\label{E10}
\end{equation}

Before we treat a real armchair carbon nanotube, let us first obtain the results for the case where there is only one subband crossing the Fermi level, $i=1$ in Eq. (7). At zero temperature, the polarization is given by the integral
\begin{equation}
P_i(q,\omega)={1\over \pi}\int ({1\over E_{i,k}-E_{i,k+q}+\hbar \omega}+{1\over E_{i,k}-E_{i,k-q}-\hbar \omega})dk,\label{E11}
\end{equation}
where the range of integration is $[-k_F,k_F]$. We obtain
\begin{eqnarray}
P_1(q,\omega)&=&{2\over a\pi E_\omega}\{ln{[\hbar\omega tan(k_Fa/4+qa/8)]^2-[4\gamma_0sin(qa/4)+E_\omega]^2\over [\hbar\omega tan(k_Fa/4+qa/8)]^2-[4\gamma_0sin(qa/4)-E_\omega]^2}\hfil\nonumber\\
&+& ln{[\hbar\omega tan(k_Fa/4-qa/8)]^2-[4\gamma_0sin(qa/4)-E_\omega]^2\over [\hbar\omega tan(k_Fa/4-qa/8)]^2-[4\gamma_0sin(qa/4)+E_\omega]^2}\},\label{E12}
\end{eqnarray}
where $E_\omega=\sqrt{[4\gamma_0sin(qa/4)]^2-(\hbar\omega)^2}$. Note that $E_\omega$ is allowed to be imaginary. When $q$ is small, $tan(k_Fa/4\pm qa/8)=tan(\pi/6\pm qa/8)\approx 1/\sqrt{3}\pm qa/6$.  Doing Taylor series expansion to lowest order in Eq. (12) gives
\begin{equation}
P_1(q,\omega)={4\gamma_0(qa)^2\over a\pi\sqrt{3}[{4\over 3}(\hbar\omega)^2-(\gamma_0qa)^2]}.\label{E13}
\end{equation}
To see how Eq. (13) is derived, note that when $q$ is small the first log term in Eq.(12) is approximately $(2/3\sqrt{3})(\hbar\omega)^2qa/[2(4\gamma_0sin(qa/4))^2-(4/3)(\hbar\omega)^2-8\gamma_0sin(qa/4)E_\omega]$, while the second log term is approximately
$-(2/3\sqrt{3})(\hbar\omega)^2qa/[2(4\gamma_0sin(qa/4))^2-(4/3)(\hbar\omega)^2+8\gamma_0sin(qa/4)E_\omega]$. Also the product of the denominators in these two expressions is exactly equal to $(4/3)(\hbar\omega)^2[(4/3)(\hbar\omega)^2-(4\gamma_0sin(qa/4))^2]$.  
From Eqs. (4) and (13), we can solve for the plasmon energy and obtain 
\begin{equation}
\hbar\omega=q{\sqrt{3}\over 2}\gamma_0a\sqrt{1+{4V_q\over \pi\sqrt{3}\gamma_0a}}.\label{E14}
\end{equation}
We note that Eq. (14) is identical to Eq. (1) obtained from Tomonaga-Luttinger models if the equivalence in Eq. (10) is taken into account. 

Of course, in real armchair carbon nanotubes, we have to take into account both subbands in Eq. (7). For $i=2$, the calculation of the polarization using Eq. (11) requires special care at the Brillouin  zone boundaries. As we can see from Fig. 1, at zero temperature the  $i=2$ subband is occupied by electrons in two segments of the subband, $[-\pi/a, -k_F]$ and $[k_F,\pi/a]$. However, if we let the integral of Eq. (11) to be over these two ranges, we get the unphysical result that the plasmon can exist only for strong and weak electron-electron interactions, but not for intermediate interaction strength. The cause of the problem is that in Eq. (11), we not only have subband energy $E_{i,k}$, but also $E_{i,k+q}$ and $E_{i,k-q}$, and the choice of the ranges $[-\pi/a, -k_F]$ and $[k_F,\pi/a]$ did not ensure that $k\pm q$ ($q>0$) be confined within the Brillouin zone.  The proper integration ranges for Eq. (11) are $[-\pi/a, -k_F]$ and $[k_F,\pi/a-q]$ for the first integrand, and $[-\pi/a+q, -k_F]$ and $[k_F,\pi/a]$ for the second integrand. With the proper integration ranges, we find the two subbands to have the same polarization, $P_2(q,\omega )= P_1(q,\omega )$. 

When $q$ is small, we can write the total polarization as
\begin{equation}
P(q,\omega)= P_1(q,\omega)+P_2(q,\omega)= {2\sqrt{3}\over \pi a\gamma_0 (x-3/4)},\label{E15}
\end{equation}
where $x=(\hbar\omega/\gamma_0qa)^2$. 
Eq. (4) then reads $1-\beta \sqrt{3}/(x-3/4)=0$, where $\beta =2V_q/\pi \gamma_0 a$, and we obtain the solution,
\begin{equation}
\hbar\omega= q{\sqrt{3}\over 2}\gamma_0a \sqrt{1+{4\beta\over \sqrt{3}}}=q{\sqrt{3}\over 2}\gamma_0a\sqrt{1+{8V_q\over \pi\sqrt{3}\gamma_0a}}.\label{E16}
\end{equation}
If we compare the result in Eq. (16) with that of Eq. (14), we see that the coefficient in front of $V_q$ has been doubled. This is because now we have two subbands, while in deriving Eq. (14) we assumed there is only one subband. More generally, if there are N (partially) occupied, symmetric subbands with degenerate Fermi wavevectors and the same bandwidth, we should have the plasmon energy
\begin{equation}
\hbar\omega=q\hbar v_F\sqrt{1+{2NV_q\over \pi \hbar v_F}}.\label{E17}
\end{equation}
The factor of 2 in front of $N$ can be traced to spin degeneracy. 
If the subbands are asymmetric and each crosses the Fermi level only once, then $N$ in Eq. (17) should be replaced by $N/2$. 

From Eqs. (9) and (16), we find the Luttinger parameter $g$ for an armchair carbon nanotube to be
\begin{equation}
g= 1/ \sqrt{1+{4\beta\over \sqrt{3}}}.\label{E18}
\end{equation}

For zigzag carbon nanotubes, 
$k_y$ is in the circumferential direction and can only take on quantized values $k_y=\nu 2\pi/na$, with $\nu =0,1,2,...,n$. For $\nu =2n/3$ (remember that $n$ is a multiple of 3 for metallic zigzag tubes), we obtain from Eq. (6) the energy dispersions of the subbands crossing the Fermi level,
\begin{equation}
E_{i,k}=\pm 2\gamma_0 sin({\sqrt{3}k_xa\over 4}).\label{E19}
\end{equation}
Figure 2 shows the dispersions of the subbands. 
For the $i=1$ subband, the proper integration ranges for Eq. (11) are $[0,\pi/\sqrt{3}a-q]$ for the first integrand, and $[0,\pi/\sqrt{3}a]$ for the second integrand. 
For the $i=2$ subband, the proper integration ranges are 
$[-\pi/\sqrt{3}a,0]$ for the first integrand, and $[-\pi/\sqrt{3}a+q,0]$ for the second integrand. 
Unlike the armchair tubes where the subbands crossing the Fermi level are symmetric, non-degenerate A bands, for zigzag tubes the subbands crossing the Fermi level are asymmetric, doubly degenerate E bands.$^{11}$  Taking into account the degeneracy, 
the total polarization is
\begin{eqnarray}
P(q,\omega)&=&{16\over a\sqrt{3}\pi E_\omega}[ln{\hbar\omega +b - E_\omega tan(\sqrt{3}qa/16) \over \hbar\omega +b + E_\omega tan(\sqrt{3}qa/16)}\hfil\nonumber\\
&+& ln{\hbar\omega -b + E_\omega tan(\sqrt{3}qa/16) \over \hbar\omega -b - E_\omega tan(\sqrt{3}qa/16)}],\label{E20}
\end{eqnarray}
where $b=4\gamma_0 sin(\sqrt{3}qa/ 8)$, and $E_\omega=\sqrt{b^2-(\hbar\omega )^2}$. For $q\rightarrow 0$, we have 
\begin{equation}
P(q,\omega) \rightarrow {8\gamma_0(qa)^2\over a\pi\sqrt{3}[{4\over 3}(\hbar\omega)^2-(\gamma_0qa)^2]}.\label{E21}
\end{equation}
From Eqs. (21) and (4), we find that the Luttinger parameter $g$ for zigzag carbon nanotubes is the same as that for armchair carbon nanotubes. 

The zigzag tube has four subbands crossing the Fermi level, but each subband crosses the Fermi level only once, unlike the armchair tube case where there are two subbands crossing the Fermi level but each subband crosses the Fermi level twice. As a result, for zigzag tubes, $N$ in Eq. (17) should be replaced by $N/2$. 

Chiral carbon nanotubes have four asymmetric subbands crossing the Fermi level, and each subband crosses the Fermi level only once. Hence N in Eq. (17) should also be replaced by N/2. The result is that all single-walled metallic carbon nanotubes have the same Luttinger parameter $g$.

We have also investigated the effect of changing the Fermi level by applying a gate voltage. For an armchair carbon nanotube, lowering the Fermi level would cause the two subbands to have different Fermi wavevectors. The $i=1$ subband has a smaller Fermi wavevector $k_{F1}=(2/a)arccos(0.5-E_F/2\gamma_0)$, while the $i=2$ subband has a larger Fermi wavevector $k_{F2}=(2/a)arccos(0.5+E_F/2\gamma_0)$. Replacing $k_F$ by $k_{F1}$ in Eq. (12) gives the polarization of the first subband, while replacing $k_F$ by $k_{F2}$ in the same equation gives the polarization of the second subband. We find that the plasmon energy or the Luttinger parameter $g$ depends on the Fermi energy $E_F$ very weakly.
To second order in $E_F$, we obtain
\begin{equation}
g_0/g=1+{3-8\beta^2 \over 18\beta (\sqrt{3}+4\beta )}({E_F\over \gamma_0})^2,\label{E22}
\end{equation}
where $g_0$ is the $g$ value for $E_F=0$. 
To first order in $E_F$, the $g$ value is unaffected by the lowering of $E_F$ from zero. This can be attributed to the fact that the two subbands have nearly linear dispersions near the K points. 
Data in Ref. 6 suggest that $g_0=0.185$, and $\beta=12.2$. Using these values, the above relation can also be written as
\begin{equation}
g/g_0=1+0.107({E_F\over \gamma_0})^2.\label{E23}
\end{equation}

Band structure calculations have shown that the carbon peapod based on the (10,10) carbon nanotube has $N=4$ symmetric subbands crossing the Fermi level with nearly degenerate Fermi wavevectors.$^{16}$
From Eq. (17) we expect the $g$ value for a $C_{60}$@(10,10) carbon peapod $g_p$ to be smaller than that for the (10,10) carbon nanotube $g_t$. If the four subbands of the carbon peapod have the same bandwidth, $g_p$ and $g_t$ would be related by
 \begin{equation}
1/g_p^2=2/g_t^2-1.\label{E24}
\end{equation}
However, since the calculations in Ref. 16 indicate that two of the subbands have narrower bandwidth than the others, the above relation would hold only approximately. Still,
experimental verification of this approximate relation should provide a test for the band structure calculations and reveal the connection between Luttinger liquids in the carbon nanotube and the carbon peapod.

The work presented here implies that the RPA is suitable for studying plasmons in carbon nanotube systems regardless if the system is a Fermi liquid or Luttinger liquid. Recently, a RPA theory has been developed for carbon nanotube bundles.$^{17}$ In the case of a 2D electron gas in a strong magnetic field, the RPA has been shown to be very good for wavevectors up to 2 times the inverse magnetic length, beyond which it is necessary to include exchange self-energy and ladder diagrams in a generalized RPA (GRPA), and the GRPA works well for arbitrarily large wavevectors.$^{18,19}$ We expect similar things to happen in a carbon nanotube system. In our future work we will compare the RPA and GRPA for carbon nanotube systems to determine the range of validity for RPA alone, and the significance of corrections due to exchange self-energy and ladder diagrams. 

\vfil\eject
\centerline{References}
1. F. D. M. Haldane, J. Phys. C: Solid State Phys. {\bf 14}, 2585 (1981).

2. A. R. Go\~ni {\it et al.}, Phys. Rev. Lett. {\bf 67}, 3298 (1991).

3. C. Dekker, Physics Today {\bf 52}, 22 (1999).

4. G. D. Mahan, {\it Many-Particle Physics} (Plenum, New York, 1981), section 4.4. J. Voit, Rep. Prog. Phys. 57, 977 (1995). 

5. M. Bockrath {\it et al.}, Nature {\bf 397}, 598 (1999).

6. Z. Yao, H. W. Ch. Postma, L. Balents, and C. Dekker, Nature {\bf 402}, 273 (1999).

7. C. Kane, L. Balents, and M. P. A. Fischer, Phys. Rev. Lett. {\bf 79}, 5086 (1997).

8. R. Egger and A. O. Gogolin, Phys. Rev. Lett. {\bf 79}, 5082 (1997).

9. M. P. A. Fisher and L. I. Glazman, in {\it Mesoscopic Electron Transport}, edited by L. P. Kouwenhoven, L. L. Sohn, and G. Sch\"on (Kluwer Academic, Boston, 1997), pp.331-373.

10. Q. P. Li, S. Das Sarma, and R. Joynt, Phys. Rev. B {\bf 45}, 13713 (1992).

11. P. J. F. Harris, {\it Carbon Nanotubes and Related Structures} (Cambridge University Press, Cambridge, 1999), chapter 4.

12. R. Saito, M. Fujita, G. Dresselhaus, and M. S. Dresselhaus, Appl. Phys. Lett. {\bf 60}, 2204 (1992).

13. R. Saito, M. Fujita, G. Dresselhaus, and M. S. Dresselhaus, Phys. Rev. B {\bf 46}, 1804 (1992).

14. J. W. Mintmire, B. I. Dunlap, and C. T. White, Phys. Rev. Lett. {\bf 68}, 631 (1992).

15. N. Hamada, S. I. Sawada, and A. Oshiyama, Phys. Rev. Lett. {\bf 68}, 1579 (1992).

16. S. Okada, S. Saito, and A. Oshiyama, Phys. Rev. Lett. {\bf 86}, 3835 (2001).

17. W. Que, J. Phys.: Condensed Matter {\bf 14}, 5239 (2002).

18. C. Kallin and B. I. Halperin, Phys. Rev. B {\bf 30}, 5655 (1984).

19. R. C\^ot\'e and A. H. MacDonald, Phys. Rev. B {\bf 44}, 8759 (1991).

\vfil\eject
\centerline{Figure Captions}
Figure 1. For an $(n,n)$ armchair carbon nanotube, only two symmetric, non-degenerate A bands cross the Fermi level $E_F=0$. The energy dispersions of these two subbands are given by Eq. (7). The subbands cross each other and the Fermi level at K points (Dirac points) $k_y=\pm 2\pi/3$. If the Fermi level is lowered to the horizontal dashed line by applying a gate voltage, the two subbands will have different Fermi wavevectors. 

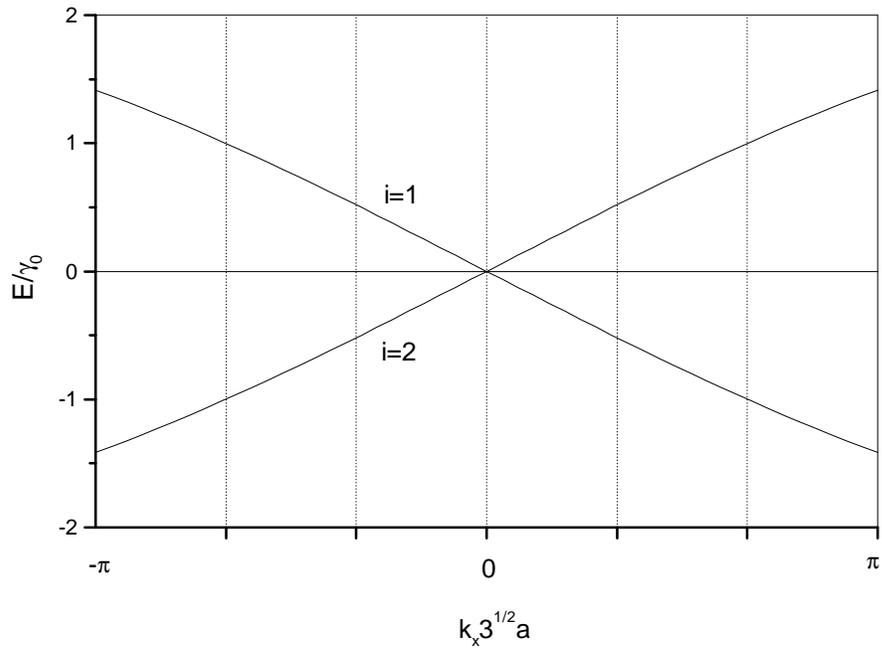
Figure 2. For a metallic zigzag carbon nanotube, asymmetric, doubly denegerate E bands cross the Fermi level. The energy dispersions of the subbands are given by Eq. (19). 

\begin{acknowledgments}
I would like to thank George Kirczenow for a discussion. 
This work was supported by the Natural Sciences and Engineering Research Council of Canada.
\end{acknowledgments}

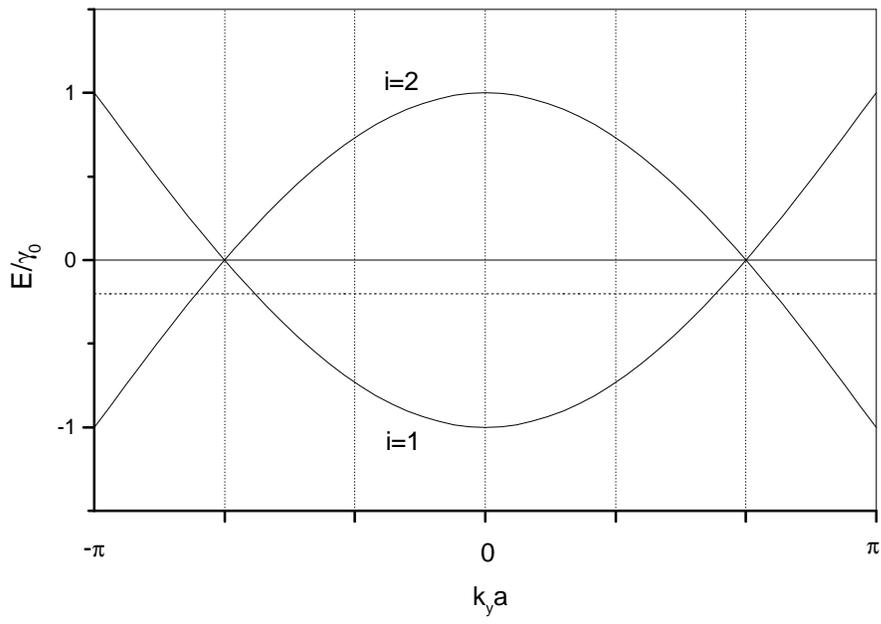
\end{document}